\documentclass[11pt]{article}
\usepackage{epsfig}
\usepackage{citesort}

\def\myendproof{{\ \vbox{\hrule\hbox{%
   \vrule height1.3ex\hskip0.8ex\vrule}\hrule }}\par}

\newtheorem{theorem}{Theorem}
\newtheorem{lemma}[theorem]{Lemma}

\newenvironment{proof}{{\it Proof. }}{\myendproof}
\newcommand{\cord}[0]{{\rm cord}}
\newcommand{\lce}[0]{{\rm lce}}
\newcommand{\dia}[0]{{\rm dia}}
\newcommand{\swallow}[1]{ }

\begin{document}

\title{A Dynamic Programming Approach to De Novo Peptide Sequencing via
Tandem Mass Spectrometry\thanks{A preliminary version appeared 
in {\em Proceedings of the 11th Annual ACM-SIAM Symposium on Discrete
  Algorithms}, pages 389--398, 2000.}}

\author{
Ting Chen
\thanks{Department of Genetics, Harvard Medical
School, Boston, MA 02115, USA.} 
\thanks{Email: tchen@salt2.med.harvard.edu. Supported
by the Lipper Foundation.}  
\and 
Ming-Yang Kao\thanks{Department of Computer Science, Yale University,
New Haven, CT 06520, USA; Email: kao@cs.yale.edu. Supported in
part by NSF Grant 9531028.}
\and
Matthew Tepel\footnotemark[1]
\and
John Rush\footnotemark[1]
\and
George M. Church\footnotemark[1]
}

\date{}

\maketitle

\begin{abstract} 

The tandem mass spectrometry fragments a large number of molecules of
the same peptide sequence into charged prefix 
and suffix subsequences,
and then measures mass/charge ratios of these ions.
The {\it de novo peptide sequencing} problem is to reconstruct
the peptide sequence from a given tandem mass spectral data of \(k\) ions.
By implicitly transforming the spectral data into an {\it NC-spectrum graph}
\(G=(V,E)\) where \(|V|=2k+2\),
we can solve this problem in \(O(|V|+|E|)\) time 
and \(O(|V|)\) space using dynamic programming.
Our approach can be further used to discover a modified 
amino acid in \(O(|V||E|)\) time and to analyze data with other types of
noise in \(O(|V||E|)\) time.
Our algorithms have been implemented and tested on actual experimental data.

\end{abstract}

\section{Introduction}

The determination of the amino acid sequence of a protein is the first
step toward solving the structure and the function of this protein. 
Conventional sequencing methods \cite{Wilkins:1997:PRN} cleave proteins into peptides
and then sequence the peptides individually using
Edman degradation or ladder sequencing by mass spectrometry
or tandem mass spectrometry \cite{McLafferty:1999:BMS}.
Among such methods, tandem mass spectrometry combined with
microcolumn liquid chromatography has been widely used as follows.
A large number of molecules of the same but unknown peptide sequence
are selected from a liquid chromatographer and a mass analyzer.
Then they are fragmented and ionized by collision-induced dissociation.
Finally all the resulting ions are measured 
by the tandem mass spectrometer for mass/charge ratios.
In the process of collision-induced dissociation,
a peptide bond at a random position is broken, 
and each molecule is fragmented into two {\it complementary} ions,
typically an N-terminal b-ion and a C-terminal y-ion.
For example, if the \(i\)th peptide bond 
of a peptide sequence of \(n\) amino acids
(\({\tt NH_2CHR_1CO-NHCHR_2CO-\cdots-NHCHR_nCOOH}\)) is broken, 
the N-terminal ion corresponds to a charged prefix subsequence
(\({\tt NH_2CHR_1CO-\cdots-NHCHR_iCO^+}\))
and the C-terminal ion corresponds a charged suffix subsequence
(\({\tt NH_2CHR_{i+1}CO-\cdots-NHCHR_nCOOH+H^+}\)).
This process fragments a large number of molecules 
of the same peptide sequence,
and therefore the resulting ions contain almost all possible 
prefix subsequences and suffix subsequences,
and display a spectrum in the tandem mass spectrometer.
All these prefix (or suffix) subsequences form a sequence
ladder where two adjacent sequences differ by one amino acid.
In the tandem mass spectrum, each ion appears at the position
of its mass because it carries a +1 charge.

\begin{figure}
\centerline{
\psfig{figure=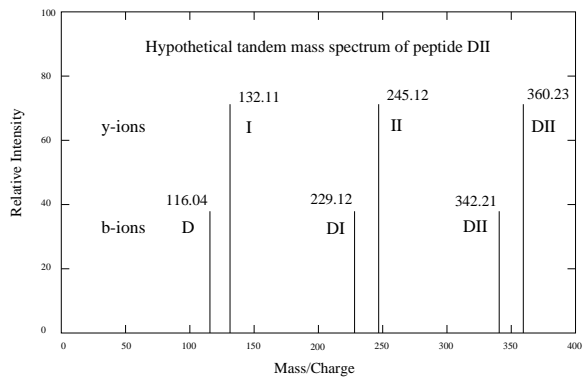,angle=270,width=3in}}
\caption{
Hypothetical tandem mass spectrum of peptide DII.
\label{fig-ms2}}
\end{figure}

Figure \ref{fig-ms2} shows all the ions 
of the peptide {\tt DII} in a hypothetical tandem mass spectrum.
The interpretation of a real tandem mass spectrum has to deal
with the following two factors:
(1) some ions may be lost in the experiments and
the corresponding mass peaks disappear in the spectrum;
(2) it is unknown whether a mass peak corresponds to
a prefix or a suffix subsequence.
The {\it de novo peptide sequencing problem} takes an input of 
a subset of prefix and suffix masses of a target peptide sequence \(P\)
and asks for a peptide sequence \(Q\) 
such that a subset of its prefixes and suffixes
gives the same input masses. Note that as expected,
\(Q\) may or may not be the same as \(P\), 
depending on the input data and the quality.

In practice, other factors can also affect a tandem mass spectrum.
An ion may display two or three different mass peaks 
because of the distribution of two isotopic carbons, 
\({\tt C^{12}}\) and \({\tt C^{13}}\), 
in the molecules.
An ion may lose a water or an ammonia molecule and displays
a different mass peak from its normal one.
An amino acid at some unknown location of the peptide sequence 
is modified and the mass is changed. 
This modification appears in every molecule of this peptide,
and all the ions containing the modified amino acid
display different mass peaks from the unmodified ions.
Finding the modified amino acid
is of great interest to biologists 
because the modification is usually associated with protein functions.

Several computer programs have been designed to
interpret the tandem mass spectral data.
A popular approach \cite{Eng:1994:ACT} is to correlate peptide sequences 
in a protein database with the tandem mass spectrum.
Peptide sequences in the database are converted into hypothetical
tandem mass spectra, which are matched against 
the target spectrum using some correlation functions,
and the sequences with top scores are reported.
This approach gives an accurate identification,
but cannot handle the peptides that are not in the database.
Also, it does not scale up very well with the length of a protein
and the size of a protein database 
because the number of peptides for a protein grows quadratically
with the length of the protein.
Pruning techniques have been applied to screen the peptides
before matching but at the cost of reduced accuracy.

An alternative approach \cite{Dancik:1999:NPS} is
{\it de novo peptide sequencing}.
The peptide sequences are extracted from the spectral data
before they are validated in the database.
First, the spectral data is transformed to a directed acyclic graph,
called a {\it spectrum graph}, where (1) a node corresponds 
to a mass peak and an edge, labeled by some amino acids, 
connects two nodes differed by the total mass of 
the amino acids in the label;
(2) a mass peak is transformed into several nodes in the graph,
and each node represents a possible prefix subsequence (ion) 
for the peak.
Then, an algorithm is called to find a longest or highest-scoring
path in the graph. 
The concatenation of edge labels in the path 
gives one or multiple candidate peptide sequences.
However, the well-known algorithms \cite{Cormen:1990:IA} for 
finding the longest path tend to include multiple nodes associated 
with the same mass peak. 
This interprets a mass peak with multiple ions of a peptide sequence, 
which is rare in practice.
This paper provides efficient sequencing algorithms
for a general interpretation of the data by
restricting a path to contain at most one node for each mass peak.

For this purpose, we introduce the notion of an {\it NC-spectrum graph} 
\(G=(V,E)\) for a given tandem mass spectrum, where \(E=2k+2\)
and \(k\) is the number of mass peaks in the spectrum.
In conjunction with this graph, 
we develop a dynamic programming approach 
to obtain the following results for previously open problems:
\begin{itemize}
\item
The de novo peptide sequencing problem can be solved
in \(O(|V|+|E|)\) time and \(O(|V|)\) space for clean spectral data,
and in \(O(|V||E|)\) time and \(O(|V|^2)\) space for noisy data.
\item
A modified amino acid can be found in \(O(|V||E|)\) time.
\end{itemize}

Our paper is organized as follows.
Section \ref{section-problem} formally defines 
the NC-spectrum graph and the peptide sequencing problem.
Section \ref{section-algorithm} describes 
the dynamic programming algorithms.
Section \ref{section-application} refines the algorithms
for the data with a modified amino acid and other types of noise.
Section \ref{section-experiment} reports the 
implementation and testing of our algorithms on experimental data.
Section \ref{section-further} mentions further research.

\section{Spectrum graphs and the peptide sequencing problem}
\label{section-problem}

Given the mass \(W\) of a target peptide sequence \(P\),
\(k\) ions \(I_1, \ldots, I_k\) of \(P\), and the masses
\(w_1, \ldots, w_k\) of these ions, we create the
{\it NC-spectrum graph} \(G=(V,E)\) as follows.

For each \(I_j\), it is unknown whether it is an N-terminal ion 
or a C-terminal ion.
If \(I_j\) is a C-terminal ion, it has a complementary N-terminal ion, 
denoted as \(I^c_j\), with a mass of \(W-w_j\).
Therefore, we create two complementary nodes $N_j$ and $C_j$
to represent \(I_j\) and \(I^c_j\), one of which must be 
an N-terminal ion.
We also create two auxiliary nodes $N_0$ and $C_0$ to
represent the zero-length and full-length N-terminal ions of \(P\).
Let \(V = \{N_0, N_1, ..., N_k, C_0, C_1, ..., C_k\}\).
Each node \(x\in V\), is placed at a real line,
and its coordinate $\cord(x)$ is the total mass of its
amino acids, i.e.,
\[ \cord(x) =
\left \{
\begin{array}{lll}
0        & x=N_0; &\\
W-18     & x=C_0; &\\
w_j-1    & x=N_j &\mbox{ for } j=1,\ldots,k;\\
W-w_j    & x=C_j &\mbox{ for } j=1,\ldots,k.\\
\end{array} 
\right.
\]
This coordinate scheme is adopted for the following reasons.
An N-terminal b-ion has an extra Hydrogen (approximately 1 dalton), 
so \(\cord(N_j)=w_j-1\) and \(\cord(C_j)=(W-(w_j-1))-1=W-w_j\);
and the full peptide sequence of $P$ has two extra Hydrogens 
and one extra Oxygen (approximately 16 daltons), so \(\cord(C_0)=W-18\).
If \(\cord(N_i)=\cord(C_j)\) for some \(i\) and \(j\),
\(I_i\) and \(I_j\) are complementary: 
one of them corresponds to a prefix sequence and 
another corresponds to the complementary suffix sequence.
In the spectrum graph, they are transformed into 
one pair of complementary nodes.
We say that $N_j$ and $C_j$ are {\it derived} from $I_j$. 
For convenience, for $x$ and $y$ $\in V$, if $\cord(x) < \cord(y)$,
then we say $x < y$.

\begin{figure}
\centerline{
\psfig{figure=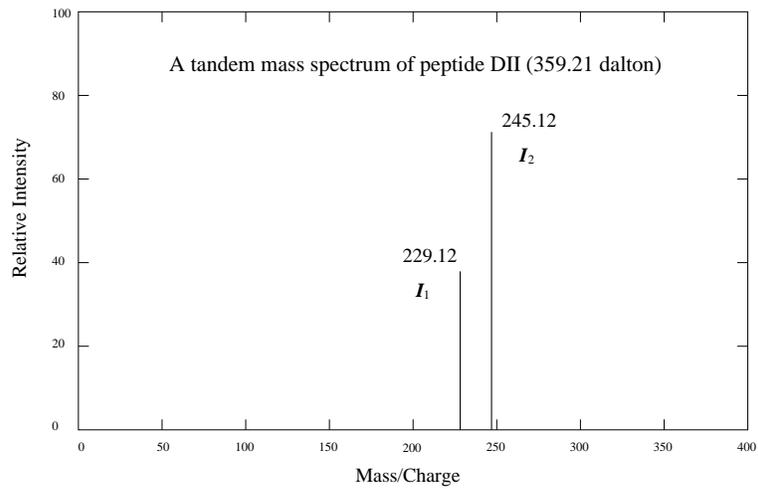,width=2.5in,angle=270}
}
\vspace{.2in}
\centerline{
\psfig{figure=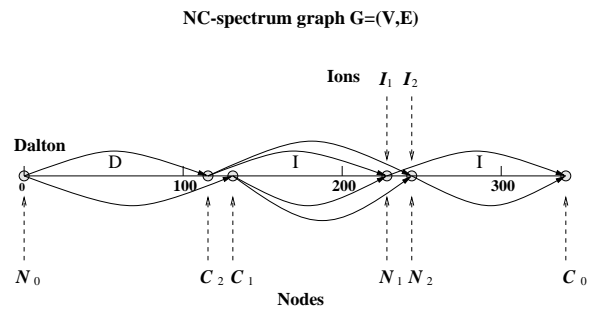,width=3in}
}
\caption{
A tandem mass spectrum and its corresponding NC-spectrum graph.
\label{fig-dii}}
\end{figure}

The edges of $G$ are specified as follows.
For $x$ and $y$ $\in V$, there is a directed edge
from $x$ to $y$, denoted by \(E(x,y)=1\),
if the following conditions are satisfied:
(1) $x$ and $y$ are not derived from the same $I_j$;
(2) $x < y$; and (3) $\cord(y)-\cord(x)$ equals the
total mass of some amino acids.
Figure \ref{fig-dii} shows a tandem mass spectrum 
and its corresponding NC-spectrum graph.

Since \(G\) is a directed graph along a line
and all edges point to the right on the real line,
we list the nodes from left to right according to
their coordinates as \(x_0,x_1,\ldots,x_k,y_k,\ldots,y_1,y_0\).

\begin{lemma}
\label{lm-definition}
The {\it peptide sequencing problem} is equivalent to the problem which,
given \(G=(V,E)\), asks for a directed path from $x_0$ to $y_0$
which contains exactly one of $x_j$ and $y_j$ for each $j>0$.
\end{lemma}

\begin{proof}
If the peptide sequence is known, we can identify 
the nodes of \(G\) corresponding to the prefix subsequences
of this peptide.
These nodes form a directed path from $x_0$ to $y_0$.
Generally the mass of a prefix subsequence does not
equal the mass of any suffix subsequence,
so the path contains exactly one of $x_j$ and $y_j$ for each $j>0$.

A satisfying directed path from $x_0$ to $y_0$ contains
all observed prefix subsequences.
If each edge on the path corresponds to one amino acid, 
we can visit the edges on the path from left to right,
and concatenate these amino acids to form a peptide sequence
that display the tandem mass spectrum.
If some edge corresponds to multiple amino acids, 
we obtain more than one peptide sequences.

Even if the mass of a prefix subsequence coincidently
equals the mass of a suffix subsequence, 
which means the directed path contains both $x_j$ and $y_j$,
we can remove either $x_j$ or $y_j$ from the path
and form a new path corresponding to multiple peptide
sequences which contain the real sequence.
\end{proof}

We call such a directed path {\it a feasible reconstruction} of $P$
or {\it a feasible solution} of $G$.
To construct \(G\), 
we use a mass array \({\cal A}\), which takes an input of mass \(m\), 
and returns 1 if \(m\) equals the total mass of some amino acids; 
and 0 otherwise.
Let \(h\) be the maximum mass under construction.
Let \(\delta\) be the measurement precision for mass. Then,

\begin{theorem}
\label{thm-problem}
Assume that we are given the maximum mass \(h\) and the mass precision \(\delta\).
\begin{enumerate}
\item \label{thm-problem-A} The mass array \({\cal A}\) can be constructed in 
\(O(\frac{h}{\delta})\) time.
\item \label{thm-problem-G} With \({\cal A}\), \(G\) can be constructed in \(O(k^2)\) time.
\end{enumerate}
\end{theorem}

\begin{proof}
These statements are proved as follows.

Statement~\ref{thm-problem-A}. 
Given a mass \(m\), \(0<m\leq h\), \({\cal A}[m]=1\) if and only if
\(m\) equals one amino acid mass,
or there exists an amino acid mass \(r<m\) 
such that \({\cal A}[m-r]=1\).
If \({\cal A}\) is computed in the order
from \({\cal A}[0]\) to \({\cal A}[\frac{h}{\delta}]\),
each entry can be determined in constant time
since there are only 20 amino acids.
The total time is \(O(\frac{h}{\delta})\).

Statement~\ref{thm-problem-G}. 
For any two nodes \(v_i\) and \(v_j\) of \(G\),
we create an edge for \(v_i\) and \(v_j\), \(E(v_i,v_j)=1\),
if and only if \(0 < \cord(v_j)-\cord(v_i)<h\) 
and \({\cal A}[\cord(v_j)-\cord(v_i)]=1\). 
There are \(O(k^2)\) pairs of nodes.
With \({\cal A}\), \(G\) can be constructed in \(O(k^2)\) time.
\end{proof}

In current practice, 
\(\delta=0.01\) dalton, and 
\(h=400\) daltons, roughly the total mass of four amino acids.
The efficiency of our algorithm will allow biologists 
to consider much larger \(h\) and much smaller \(\delta\).

\section{Algorithms for peptide sequencing}
\label{section-algorithm}

\subsection{Dynamic programming}

We list the nodes of \(G\) from left to right as
\(x_0,x_1,\ldots,x_k,y_k,\ldots,y_1,y_0\).
Let \(M(i,j)\) be a two-dimension table with \(0 \leq i,j \leq k\).
Let \(M(i,j)=1\) if and only if in \(G\), there is a path \(L\) 
from \(x_0\) to \(x_i\) and a path \(R\) from \(y_j\) to \(y_0\), 
such that \(L \cup R\) contains exactly one of \(x_p\) and \(y_p\) 
for every \(p \in [0,i] \cup [0,j]\).
Let \(M(i,j)=0\) otherwise.

\begin{tabbing}
{\bf Algorithm Compute-M}\((G)\)\\
1. Initialize \(M(0,0)=1\) and \(M(i,j)=0\) for all \(i \neq 0\) or \(j \neq 0\);\\
2. Compute \(M(1,0)\) and \(M(0,1)\);\\
3. For \= \(j=2\) to \(k\)\\
4. \> For \= \(i=0\) to \(j-2\)\\
\> (a)\> if \(M(i,j-1)=1\) and \(E(x_i,x_j)=1\), then \(M(j,j-1)=1\);\\
\> (b)\> if \(M(i,j-1)=1\) and \(E(y_j,y_{j-1})=1\), then \(M(i,j)=1\);\\
\> (c)\> if \(M(j-1,i)=1\) and \(E(x_{j-1},x_j)=1\), then \(M(j,i)=1\);\\
\> (d)\> if \(M(j-1,i)=1\) and \(E(y_j,y_i)=1\), then \(M(j-1,j)=1\).
\end{tabbing}

\swallow{
\\\\
\noindent
{\bf Algorithm Compute-M}\((G)\)
\begin{enumerate}
\item \label{alg-m-ini} Initialize \(M(0,0)=1\) and \(M(i,j)=0\) 
			for all \(i \neq 0\) or \(j \neq 0\);
\item \label{alg-m-cal} Compute \(M(1,0)\) and \(M(0,1)\);
\item \label{alg-m-j} For \(j=2\) to \(k\)
\item \label{alg-m-i}\ \ \ \ For \(i=0\) to \(j-2\)
\begin{enumerate}
\item \label{alg-m-i-1}
if \(M(i,j-1)=1\) and \(E(x_i,x_j)=1\), then \(M(j,j-1)=1\);
\item \label{alg-m-i-2}
if \(M(i,j-1)=1\) and \(E(y_j,y_{j-1})=1\), then \(M(i,j)=1\);
\item \label{alg-m-i-3}
if \(M(j-1,i)=1\) and \(E(x_{j-1},x_j)=1\), then \(M(j,i)=1\);
\item \label{alg-m-i-4}
if \(M(j-1,i)=1\) and \(E(y_j,y_i)=1\), then \(M(j-1,j)=1\).
\end{enumerate}
\end{enumerate}
}

\begin{lemma}
\label{lm-compute-m}
Given \(G(V,E)\),
{\rm Algorithm Compute-M} computes the table \(M\) in \(O(|V|^2)\) time.
\end{lemma}
\begin{proof}
Let \(L\) and \(R\) be the paths that correspond to \(M(i,j)=1\).
If \(i<j\), by definition, after removing node \(y_j\) from \(R\),
\(L \cup R-\{y_j\}\) contains
exactly one of \(x_q\) and \(y_q\) for all \(1 \leq q \leq j-1\).
If \((y_j,y_p) \in R\), then \(M(i,p)=1\),
and either \(p=j-1\) or \(i=j-1\), which corresponds to 
Step 4(b) or 4(d) respectively in the algorithm,
because either \(x_{j-1}\) or \(y_{j-1}\), but not both, is in \(L \cup R\).
A similar analysis holds for the cases of Step 4(a) or 4(c).
The loop at Step 3
uses previously computed
\(M(0,j-1),\ldots,M(j-1,j-1)\) and \(M(j-1,0),\ldots,M(j-1,j-1)\)
to fill up \(M(0,j),\ldots,M(j,j)\) and \(M(j,0),\ldots,M(j,j)\).
Thus the algorithm computes \(M\) correctly.
Note that \(|V|=2k+2\) and Steps 4(a), 4(b), 4(c), and 4(d)
take \(O(1)\) time, and thus the total time is \(O(|V|^2)\).
\end{proof}

\begin{theorem}
\label{thm-feasible}
The following statements hold.
\begin{enumerate}
\item \label{thm-feasible-GM}
Given \(G=(V,E)\) and \(M\), 
a feasible solution of \(G\) can be found in \(O(|V|)\) time.
\item \label{thm-feasible-G}
Given \(G=(V,E)\),
a feasible solution of \(G\) can be found in
 \(O(|V|^2)\) time and \(O(|V|^2)\) space.
\item \label{thm-feasible-all}
Given \(G=(V,E)\),
all feasible solutions of \(G\) can be found in
\(O(|V|^2+n|V|)\) time and \(O(|V|^2+n|V|)\) space,
where \(n\) is the number of solutions.
\end{enumerate}
\end{theorem}

\begin{proof} These statements are proved as follows.

Statement~\ref{thm-feasible-GM}.
Note that \(|V|=2k+2\).
Without loss of generality, assume that a feasible solution \(S\) contain
node \(x_k\). 
Then there exists some \(j<k\), such that \((x_k,y_j)\) is an edge in \(S\)
and \(M(k,j)=1\). 
Therefore, we search the non-zero entries in the last row of \(M\)
and find a \(j\) that satisfies both \(M(k,j)=1\) and \(E(x_k,y_j)=1\).
This takes \(O(|V|)\) time.
With \(M(k,j)=1\), we backtrack \(M\) to search the next edge of \(S\) as follows.
If \(j=k-1\), the search starts from \(i={k-2}\) to \({0}\) until
both \(E(x_i,x_k)=1\) and \(M(i,j)=1\) are satisfied;
otherwise \(j<k-1\), and then \(E(x_{k-1},x_k)=1\) and \(M(k-1,j)=1\).
We repeat this process to find every edge of \(S\).
The process visits every node of \(G\) at most once
in the order from \(x_k\) to \(x_0\) and from \(y_k\) to \(y_0\).
The total cost is \(O(|V|)\) time.

Statement~\ref{thm-feasible-G}.
We compute \(M\) by means of Lemma \ref{lm-compute-m} and
find a feasible solution by means of Statement~\ref{thm-feasible-GM}.
The total cost is \(O(|V|^2)\) time and \(O(|V|^2)\) space.

Statement~\ref{thm-feasible-all}.
The proof is similar to that of Statement~\ref{thm-feasible-GM}.
We can find all the feasible solutions by backtracking \(M\),
and each feasible solution costs \(O(|V|)\) time 
and \(O(|V|)\) space.
Computing \(M\) and finding 
\(n\) solutions cost \(O(|V|^2+n|V|)\) time and  
\(O(|V|^2+n|V|)\) space in total.
\end{proof}

\subsection{An improved algorithm}

To improve the time and space complexities in Theorem \ref{thm-feasible},
we encode \(M\) into two linear arrays.
Define an edge \((x_i,y_j)\) with \(0 \leq i,j \leq k\) to be a {\it cross edge},
and an edge \((x_i,x_j)\) or \((y_j,y_i)\)
with \(0 \leq i < j \leq k\) to be an {\it inside edge}. 
Let \(\lce(z)\) be the length of the longest consecutive inside edges starting 
from node \(z\); i.e.,
\[
\left \{
\begin{array}{ll}
\lce(x_i)=j-i      & {\rm if}\ E(x_i,x_{i+1})=\ldots =E(x_{j-1},x_j)=1 \ {\rm and} \ (j=k \ {\rm or} \ E(x_j,x_{j+1})=0); \\
\lce(y_i)=i-j      & {\rm if}\ E(y_i,y_{i-1})=\ldots =E(y_{j+1},y_j)=1 \ {\rm and} \ (j=0 \ {\rm or} \ E(y_j,y_{j-1})=0). 
\end{array} 
\right.
\]
Let \(\dia(z)\) be two diagonals in \(M\), where
\[
\left \{
\begin{array}{ll}
\dia(x_j)=M(j,j-1) & {\rm for }\ 0<j\leq k;\\
\dia(y_j)=M(j-1,j) & {\rm for }\ 0<j\leq k;\\
\dia(x_0)=\dia(y_0)=1.
\end{array} 
\right.
\]

\begin{lemma}
\label{lm-compact}
Given \(\lce(\cdot)\) and \(\dia(\cdot)\), any entry of \(M\) can be computed in \(O(1)\) time.
\end{lemma}

\begin{proof}
Without loss of generality, let the \(M(i,j)\) be the entry we
want to compute where \(0\leq i<j \leq k\).
If \(i=j-1\), \(M(i,j)=\dia(y_j)\) as defined; otherwise \(i<j-1\)
and \(M(i,j)=1\) if and only if \(M(i,i+1)=1\) and
\(E(y_j,y_{j-1})=\ldots=E(y_{i+2},y_{i+1})=1\),
which is equivalent to \(\dia(y_{i+1})=1\) and \(\lce(y_j) \geq j-i-1\).
Thus both cases can be solved in \(O(1)\) time.
\end{proof}

\begin{lemma}
\label{lm-lce-dia}
Given \(G=(V,E)\), \(\lce(\cdot)\) and \(\dia(\cdot)\) can be computed in 
\(O(|V|+|E|)\) time.
\end{lemma}

\begin{proof}
We retrieve consecutive edges starting from \(y_k\), \(y_{k-1}\), \(\ldots\),
until the first \(y_p\) with \(p\leq k\) and \(E(y_{p},y_{p-1})=0\).
Then we can fill \(\lce(y_k)=k-p\), \(\lce(y_{k-1})=k-p-1\), \(\ldots\),
and \(\lce(y_p)=0\) immediately.
Next, we start a new retrieving and filling process from \(y_{p-1}\),
and repeat this until \(y_0\) is visited.
Eventually we retrieve \(O(k)\) consecutive edges.
A similar process can be applied to \(x\).
Using a common graph data structure such as a link list, 
a consecutive edge can be retrieved in constant time,
and thus \(\lce(\cdot)\) can be computed in \(O(|V|)\) time.

By definition, \(\dia(x_j)=M(j,j-1)=1\) if and only if 
there exists some \(i\) with \(0\leq i<j-1\),
\(M(i,j-1)=1\), and \(E(x_i,x_j)=1\).
If we have computed \(\dia(x_0),\ldots,\dia(x_{j-1})\) and
\(\dia(y_{j-1}),\ldots,\dia(y_0)\),
then \(M(i,j-1)\) can be computed in constant time 
by means of the proof in Lemma \ref{lm-compact}.
To find the \(x_i\) for \(E(x_i,x_j)=1\),
we can visit every inside edge that ends at \(x_j\).
Therefore the computation of \(\dia(\cdot)\) visits
every inside edge exactly once, 
and the total time is \(O(|V|+|E|)\).
\end{proof}

\begin{theorem}
\label{thm-improved}
Assume that \(G(V,E)\) is given.
\begin{enumerate}
\item \label{thm-improved-one}
A feasible solution of \(G\) can be found in \(O(|V|+|E|)\) time 
and \(O(|V|)\) space.
\item  \label{thm-improved-all}
All feasible solutions of \(G\) can be found in \(O(n|V|+|E|)\) time 
and \(O(n|V|)\) space, where \(n\) is the number of solutions.
\end{enumerate}
\end{theorem}

\begin{proof} These statements are proved as follows.

Statement~\ref{thm-improved-one}.
By Lemma \ref{lm-lce-dia}, \(\lce(\cdot)\) and \(\dia(\cdot)\) can be computed
in \(O(|V|+|E|)\) time and \(O(|V|)\) space.
By Lemma \ref{lm-compact}, the last row and the last column of \(M\)
can be reconstructed from \(\lce\) and \(\dia\) in \(O(|V|)\) time.
By Theorem \ref{thm-feasible} and Lemma \ref{lm-compact},
a feasible solution of \(G\) can be found in \(O(|E|)\) time.
Therefore, finding a feasible solution takes 
\(O(|V|+|E|)\) time and \(O(|V|)\) space.

Statement~\ref{thm-improved-all}.
The proof is similar to the proof of Statement~\ref{thm-feasible-all} 
in Theorem \ref{thm-feasible}. 
Finding an additional feasible solution takes \(O(|V|)\) time
and \(O(|V|)\) space.
Thus finding \(n\) solutions takes \(O(n|V|+|E|)\) time and \(O(n|V|)\) space.
\end{proof}

A feasible solution of \(G\) is a path of \(k+1\) nodes and \(k\) edges,
and therefore there must exist an edge between any two nodes
on the path by the edge transitive relations.
This implies that there are at least \((k+1)k/2\) 
or \(O(|V|^2)\) edges in the graph.
However, in practice, a threshold is usually set for the maximum length (mass)
of an edge, so the number of edges in \(G\) could be much smaller 
than \(O(|V|^2)\) and may actually equal \(O(|V|)\) sometimes.

\section{Algorithms for noisy data}
\label{section-application}

\subsection{Amino acid modification}

Amino acid modifications are related to protein functions.
For example, some proteins are active when phosphorylated
but inactive when dephosphorylated.
Although there are a few hundred known modifications,
a peptide rarely has two or more modified amino acids.
This section discusses how to find the position
of a modified amino acid from a tandem mass spectral data.
We assume that the modified mass is unknown and
is not equal to the total mass
of any number of amino acids; otherwise,
it is information-theoretically impossible to detect
an amino acid modification from tandem mass spectral data.

\begin{lemma}
\label{lm-modification}
The amino acid modification problem is equivalent to
the problem which, given \(G=(V,E)\), asks for
two nodes \(v_i\) and \(v_j\), such that \(E(v_i,v_j)=0\) 
but adding the edge \((v_i,v_j)\) to \(G\) creates a 
feasible solution that contains this edge.
\end{lemma}

\begin{proof}
Similar to Lemma \ref{lm-definition}.
\end{proof}

\swallow{
we define the {\it amino acid modification problem}:
given a NC-spectrum graph \(G\), ask to find two node \(v_i\) and \(v_j\),
such that \(E(v_i,v_j)=0\) but adding the edge \((v_i,v_j)\)
to \(G\) creates a feasible solution that contains this edge.
}

Let \(G=(V,E)\) be an NC-spectrum graph with nodes
from left to right as \(x_0,\ldots,x_k,y_k,\ldots,y_0\).
Let \(N(i,j)\) be a two-dimension table with \(0 \leq i,j \leq k\),
where \(N(i,j)=1\) if and only if there is a path from \(x_i\) to
\(y_j\) which contains exactly one of \(x_p\) and \(y_p\) 
for every \(p \in [i,k] \cup [j,k]\).
Let \(N(i,j)=0\) otherwise.
\swallow{
\\\\
\noindent
{\bf Algorithm Compute-N}\((G)\)
\begin{enumerate}
\item \label{alg-n-ini} Initialize \(N(i,j)=0\) for all \(i\) and \(j\);
\item \label{alg-n-cal} Compute \(N(k,k-1)\) and \(N(k-1,k)\);
\item \label{alg-n-j} For \(j=k-2\) to \(0\)
\item \label{alg-n-i}\ \ \ \ For \(i=k\) to \(j+2\)
\begin{enumerate}
\item \label{alg-n-i-1}
if \(N(i,j+1)=1\) and \(E(x_j,x_i)=1\), then \(N(j,j+1)=1\);
\item \label{alg-n-i-2}
if \(N(i,j+1)=1\) and \(E(y_{j+1},y_j)=1\), then \(N(i,j)=1\);
\item \label{alg-n-i-3}
if \(N(j+1,i)=1\) and \(E(x_j,x_{j+1})=1\), then \(N(j,i)=1\);
\item \label{alg-n-i-4}
if \(N(j+1,i)=1\) and \(E(y_i,y_{j+1})=1\), then \(N(j+1,j)=1\).
\end{enumerate}
\end{enumerate}
}

\begin{tabbing}
{\bf Algorithm Compute-N}\((G)\)\\
1. Initialize \(N(i,j)=0\) for all \(i\) and \(j\);\\
2. Compute \(N(k,k-1)\) and \(N(k-1,k)\);\\
3. For \= \(j=k-2\) to \(0\)\\
4.  \>  For \= \(i=k\) to \(j+2\)\\
\>  (a)\>  if \(N(i,j+1)=1\) and \(E(x_j,x_i)=1\), then \(N(j,j+1)=1\);\\
\>  (b)\>  if \(N(i,j+1)=1\) and \(E(y_{j+1},y_j)=1\), then \(N(i,j)=1\);\\
\>  (c)\>  if \(N(j+1,i)=1\) and \(E(x_j,x_{j+1})=1\), then \(N(j,i)=1\);\\
\>  (d)\>  if \(N(j+1,i)=1\) and \(E(y_i,y_{j+1})=1\), then \(N(j+1,j)=1\).
\end{tabbing}

\begin{lemma}
\label{lm-compute-n}
Given \(G=(V,E)\),
{\rm Algorithm Compute-N} computes the table \(N\) in \(O(|V|^2)\) time.
\end{lemma}

\begin{proof}
Similar to Lemma \ref{lm-compute-m}.
\end{proof}

\begin{theorem}
\label{thm-modification}
Given \(G=(V,E)\) which contains all prefix and suffix nodes, 
all possible amino acid modifications
can be found in \(O(|V||E|)\) time and \(O(|V|^2)\) space. 
\end{theorem}

\begin{proof}
Let \(M\) and \(N\) be two tables for \(G\) computed from 
Lemma \ref{lm-compute-m} and \ref{lm-compute-n}.
Without loss of generality,
let the modification be 
between two consecutive prefix nodes \(x_i\) and \(x_j\)
with \(0\leq i<j \leq k\) and \(E(x_i,x_j)=0\).
All the prefix nodes to the right of \(x_j\) 
have the same mass offset from the normal locations
because the corresponding sequences contain the modified amino acid.
By adding a new edge \((x_i,x_j)\) to \(G\),
we create a feasible solution \(S\) that contains this edge.
If \(i+1<j\), then \(y_{i+1}\in S\),
and thus \(M(i,i+1)=1\) and \(N(j,i+1)=1\).
There are \(O(k^2)\) possible combinations of \(i\) and \(j\),
and checking all of them takes \(O(|V|^2)\) time.
If \(i+1=j\), then \(S\) must contain an edge \((y_q,y_p)\) with \(q>j>i>p\),
which skips over \(y_i\) and \(y_j\).
\(S\) can be found if \(E(y_q,y_p)=1\) and \(M(i,p)=1\) and \(N(j,q)=1\).
There are at most \(O(|E|)\) edges, which can be examined in \(O(|E|)\) time.
Checking \(O(|V|)\) possible \(i+1=j\) costs \(O(|V||E|)\) time.
The total complexity is \(O(|V||E|)\) time and \(O(|V|^2)\) space.
\end{proof}

Note that the condition in Theorem \ref{thm-modification}
does not require that all ions in the spectrum are observed.
If some ions are lost but their complementary ions appear, 
\(G\) still contains all prefix and suffix nodes of the target sequence.
Furthermore, if \(G\) does not contain all prefix and suffix nodes
because of many missing ions, we can still use this algorithm to
find the modification but the result depends on the quality of
the data and the modified mass.

\subsection{Using scoring functions}

In practice, a tandem mass spectral data may contain noise such as
mass peaks of other types of ions from the same peptide,
mass peaks of ions from other peptides, and mass peaks of unknown ions.
A common way to deal with these situations is to use
a pre-defined edge scoring function \(s(\cdot)\).
With \(s\), the score of a path is the sum of the scores
of the edges on the path.
We re-define the {\it peptide sequencing problem},
which given an NC-spectrum graph \(G=(V,E)\), 
asks for a maximum score path from $x_0$ to $y_0$,
such that at most one of $x_j$ and $y_j$ for every $1 \leq j \leq k$ 
is on the path.

Let \(Q(i,j)\) be a two-dimension table with \(0 \leq i,j \leq k\).
\(Q(i,j)>0\) if and only if in \(G\), there is a path \(L\) 
from \(x_0\) to \(x_i\) and a path \(R\) from \(y_j\) to \(y_0\),
such that at most one of $x_p$ and $y_p$ is in \(L \cup R\) 
for every \(p \in [0,i] \cup [0,j]\); \(Q(i,j)=0\) otherwise.
If \(Q(i,j)>0\), 
let \(Q(i,j)\) be the maximum score among all \(L\) and \(R\) pairs.

\begin{tabbing}
{\bf Algorithm Compute-Q}\((G)\)\\
1. Initialize \(Q(0,0)=1\) and \(Q(i,j)=0\) for all \(i \neq 0\) or \(j \neq 0\);\\
2. For \= \(j=1\) to \(k\)\\
3. \>  For \= \(i=0\) to \(j-1\)\\
\> (a)\> For every \(E(y_j,y_p)=1\) and \(Q(i,p)>0\), \(Q(i,j)=\max\{Q(i,j), Q(i,p)+s(y_j,y_p)\}\);\\
\> (b)\> For every \(E(x_p,x_j)=1\) and \(Q(p,i)>0\), \(Q(j,i)=\max\{Q(j,i), Q(p,i)+s(x_p,x_j)\}\).
\end{tabbing}

\begin{lemma}
\label{lm-compute-r}
Given \(G=(V,E)\),
{\rm Algorithm Compute-Q} computes the table \(Q\) in \(O(|V||E|)\) time.
\end{lemma}

\begin{proof}
The correctness proof is similar to that for Lemma \ref{lm-compute-m}.
For every \(j\), Steps 3(a) and 3(b)
visit every edge of \(G\) at most once, so the total time is \(O(|V||E|)\).
\end{proof}

\begin{theorem}
Given \(G=(V,E)\),
a feasible solution of \(G\) can be found in \(O(|V||E|)\) time and \(O(|V|^2)\) space.
\end{theorem}

\begin{proof}
Algorithm Compute-Q computes \(Q\) in \(O(|V||E|)\) time and \(O(|V|^2)\) space.
For every \(i\) and \(j\), if \(Q(i,j)>0\) and \(E(x_i,y_j)=1\),
we compute the sum \(Q(i,j)+s(x_i,y_j)\).
Let \(Q(p,q)+s(x_p,y_q)\) be the maximum value,
and we can backtrack \(Q(p,q)\) to find all the edges of the feasible solution.
The total cost is \(O(|V||E|)\) time and \(O(|V|^2)\) space.
\end{proof}

\section{Experimental results}
\label{section-experiment}

We have presented algorithms for reconstructing peptide sequences
from a tandem mass spectral data with loss of ions.
This section reports experimental studies which focus
on cases of b-ions losing a water or ammonia molecule 
and cases of isotopic varieties for an ion.
We treat the rare occurrence 
such as y-ions losing a water or ammonia molecule,
b-ions losing two water or ammonia molecules, 
and other types of ions,
as noise and apply Algorithm Compute-Q to reconstruct peptide sequences.

Isotopic ions come from isotopic carbons of 
\({\tt C^{12}}\) and \({\tt C^{13}}\). 
An ion usually has a couple of isotopic forms,
and the mass difference between two isotopic ions 
is generally one or two daltons.
Their intensities reflect the binomial distribution
between \({\tt C^{12}}\) and \({\tt C^{13}}\).
This distribution can be used for identification.
Isotopic ions can be merged to one ion of
either the highest intensity or a new mass.

It is very common for a b-ion to lose a water or ammonia molecule.
In the construction of an NC-spectrum graph, we add three
types of edges whose lengths equal the masses of
a water molecule, amino acids minus one water, 
and amino acids plus one water respectively.
In Algorithm Compute-Q, we restrict the net number of waters
at each entry to be at most one,
since a feasible solution should have a net of zero water.
We have implemented this algorithm and tested it on the data generated by
the following process:
\begin{quote}
The Chicken Ovalbumin proteins 
were digested with trypsin in 100 mM ammonium bicarbonate buffer pH 8 for 18
hours at \(37^{\circ}C\). Then 100 \(\mu\ell\) are injected in acetonitrile
into a reverse phase HPLC interfaced 
with a Finnigan LCQ ESI-MS/MS mass spectrometer.
A 1\% to 50\% acetonitrile 0.1\%TFA linear gradient was executed over 60 minutes.
\end{quote}

\begin{figure}
\centerline{
\psfig{figure=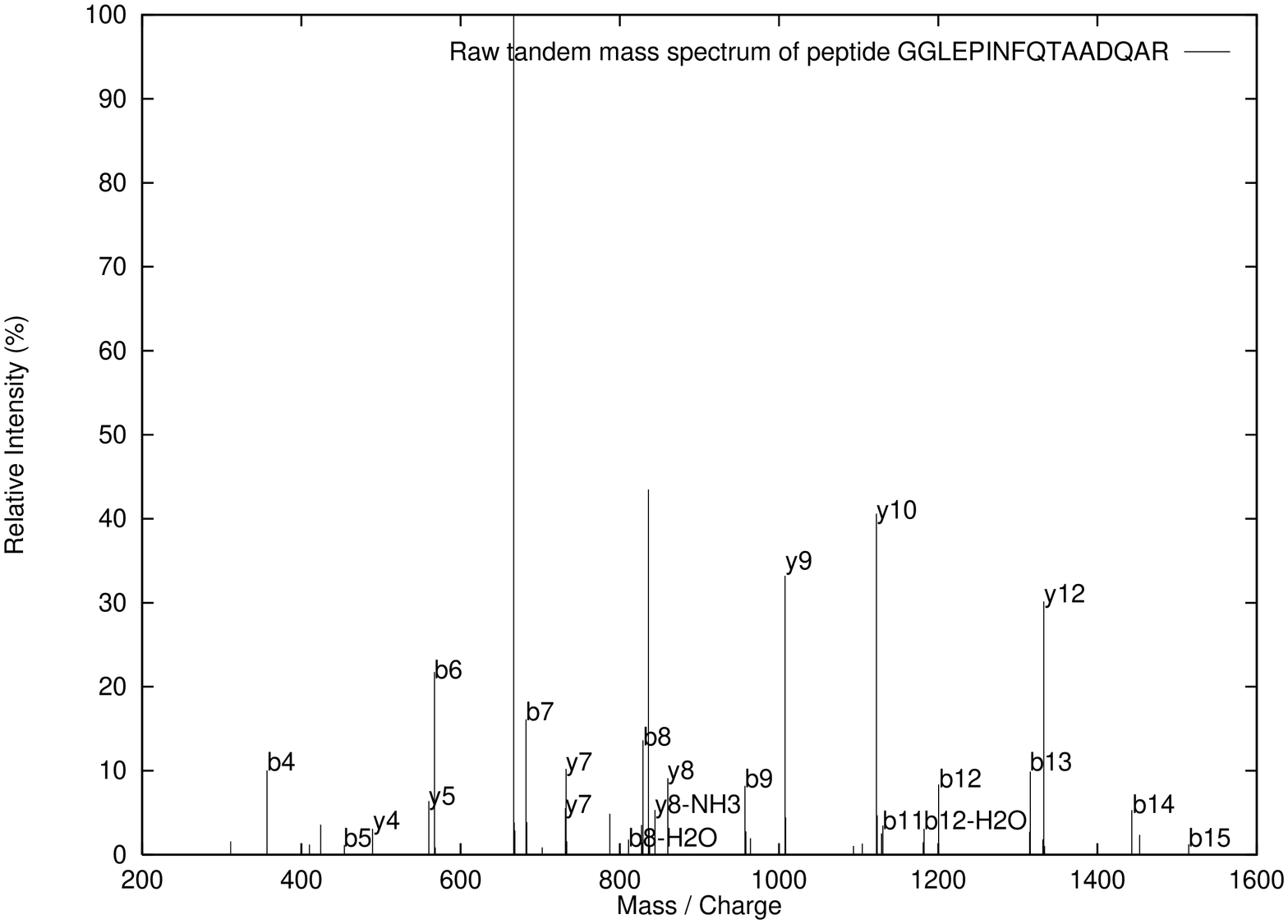,angle=270,width=2.5in}
\psfig{figure=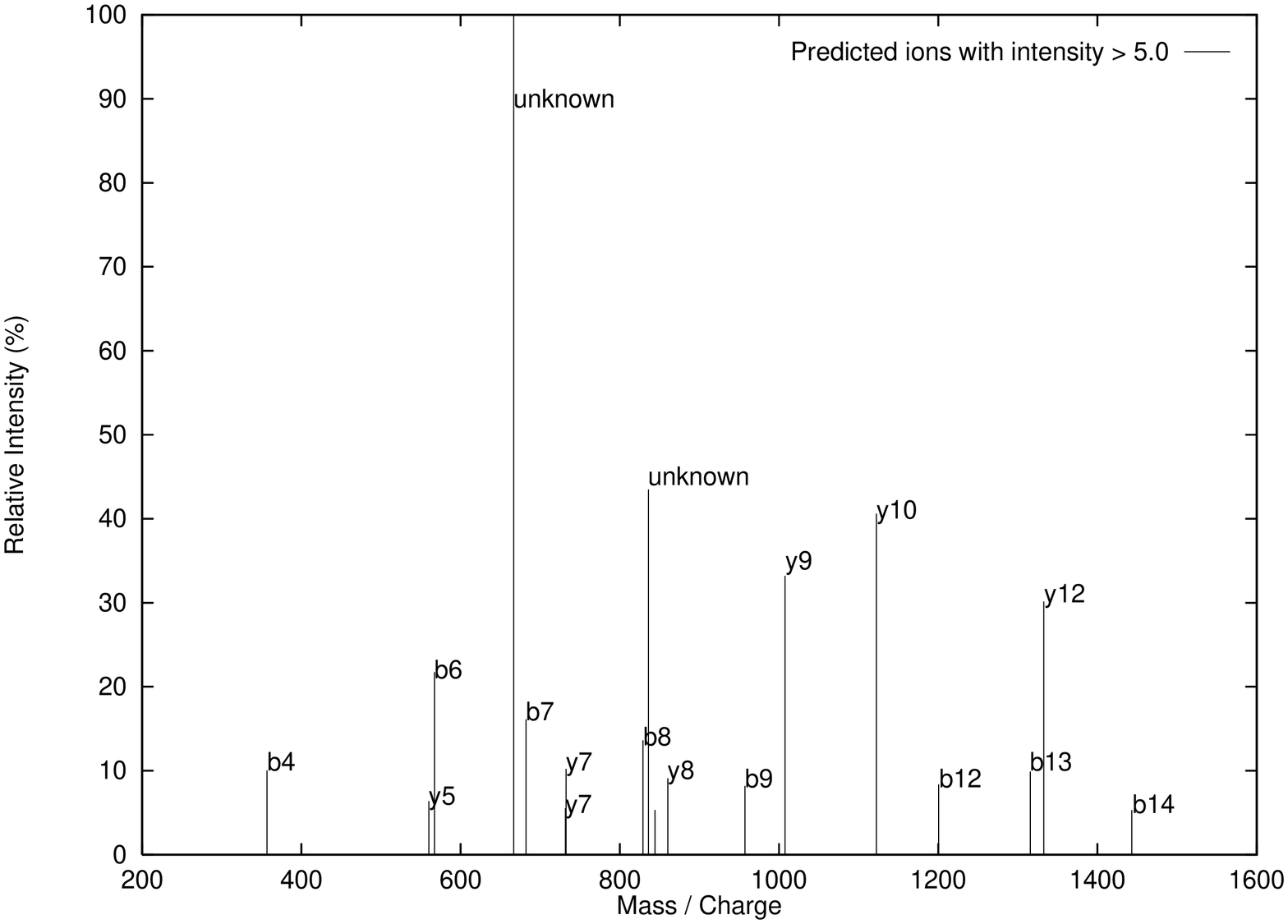,angle=270,width=2.5in}
}
\caption{
Raw tandem mass spectrum and predicted ions 
of the Chicken Ovalbumin peptide GGLEPINFQTAADQAR.
\label{fig-real}}
\end{figure}

Figure \ref{fig-real} shows one of our prediction results.
The ions labeled in the spectrum were identified successfully.
We use resolution 1.0 dalton and relative intensity threshold 5.0 in our program.
More experimental results will be shown in the full version
of this paper.

\section{Further research}
\label{section-further}
There are many open problems. Perhaps the most interesting direction would
be to consider the case of multiple peptides.

\bibliographystyle{plain}
\bibliography{all}

\end{document}